\documentclass[aps,twocolumn, prb]{revtex4}
                                                                                                                                                             
\usepackage{graphicx}
\usepackage{amssymb}
\usepackage{amsmath}
\usepackage{natbib}
\usepackage{verbatim}
\usepackage{mathrsfs}
\usepackage{times}
\usepackage{epstopdf} 
\usepackage{psfrag}                                                                                                                                                           
\begin{document}

\newcommand{\vect}[1] {\vec{#1}} %%% whatever typesetting we want to use for vectors
\newcommand{\unit}[1] {\hat{#1}}  %%% typesetting for unit vectors
\newcommand{\Hamil} {{\mathbf{H}}} %%% typesetting for Hamiltonian symbol
\newcommand{\spin} {{\mathbf{S}}} %%% typesetting for Spin symbol
\newcommand{\dotp}[2] { {\vect #1} \cdot {\vect #2}} %%%% typesetting for a dot product
\newcommand{\en} {\chi} %%%%% typesetting for the dispersion symbol
\newcommand{\ti}[1] {\tilde{#1}} %%%%% typesetting for putting tilde on #1
                                                                                                                                                         
\title{Effect of Dirac Spinons on ARPES signatures of Herbertsmithite}
\author{Sumiran Pujari}
\affiliation{Department of Physics,
Cornell University, Ithaca, New York 14853-2501.}
\author{Michael Lawler}
\affiliation{Department of Physics,
Cornell University, Ithaca, New York 14853-2501}
\affiliation{Department of physics, Binghamton University, Vestal NY 13850.}

\email{sp384@cornell.edu}
                                                                                                                                                             
\begin{abstract}
The spinon continues to be an elusive elementary excitation of frustrated antiferromagnets.
To solidify evidence for its existence, we address the question of what will be the 
Angle Resolved Photoemission Spectroscopy (ARPES) signatures of single crystal 
samples of Herbertsmithite assuming it is described by the Dirac spin liquid state. 
In particular, we show that the electron spectral function will have a linear in energy
dependence near specific wave vectors and that this dependence is expected even 
after fluctuations to the mean field values are taken into account. Observation of this 
unique signature in ARPES will provide very strong evidence for the 
existence of spinons in greater than one dimension.
\end{abstract}

\maketitle

\section{Introduction}
\begin{comment}Frustrated quantum magnetism is a topic of continuing interest since the proposal of a
spin liquid ground state by Anderson in the 70s. Since then, the possibility of frustration
(in geometric or non-geometric guises) melting the magnetic order continues to be intensely
explored by many theoreticians. Many exotic ground states have been proposed in the process
and many more will be. The main hindrance in settling these questions with full finality
is always our inability to fully solve the model hamiltonians(except in one dimension where
we can use Bosonization and Bethe Ansatz techniques to fully solve for the Eigenspectrum)
in the regime of interest. We have to resort to various approximations to make progress 
and the resulting conclusions always come with a caveat of their being not applicable 
to the regime of interest. This is where experiment(numerical and physical) help us.\end{comment}

\begin{comment} % outline of Sumiran's introduction
A. frustrated magnets are interesting
B. Herbertsmithite is interesting and appears to be a gapless spin liquid
C. History of kagome antiferromagnetic Heisenberg model
D. Usefulness of ARPES 
E. Dirac spin liquid state
F. Organization of the paper
\end{comment}

\begin{comment}
A. What is the problem and why is it hard
B. How have people studied this problem and what were past limitations.
C. What is our new approach and how has it succeeded
D. Why should we care (what are the broad implications) --> Discussion section

\end{comment}

%\textbf{ highlight our main problem: How can we find evidence for the spinon? }

Frustrated quantum magnetism is of active current interest because of the possibility
of supporting exotic ground states and excitations. 
One of the 
central issue that is driving the field of frustrated magnetism is the search for 
such exotic objects in Nature. 
A prime example of an exotic excitation is a Spinon - a spin-1/2 fermion that is charge neutral
- which is also the main object of our study. 
The existence of a spin-1/2 fermion particle 
that can be excited out of a ground state of ordinary spins would indeed be remarkable;
thus, it is important and appropriate to flesh out what would consitute 
as concrete evidence of the existence of the spinon.

%\textbf{ Search for spinons in 1D organometalic compounds, cuprates and NiGa2S4 via ARPES}

In one dimensional systems, there is substantial evidence for the existence of 
spin 1/2 solitons \cite{1DSpinons}, though these are different 
types of excitations than the 
spin 1/2 spinon proposed in frustrated 
magnets\cite{Mudry}. One remarkable 
source of this evidence is from ARPES experiments 
performed on SrCuO$_2$\cite{KimUchida}, 
a tool which has been successfully applied to a number of 
Mott insulators\cite{ARPESreview}. Even though ARPES experiments require single 
crystals in order to have good momentum resolution, they are 
feasible with smaller crystals than what is required for neutron 
diffraction studies. ARPES may also thus provide an additional avenue to 
falsify any of the various ground state proposals 
in the frustrated antiferromagnets. 

% Organization of paper: here we guide different readers through the paper and help, for example, expermentalists choose which sections to focus on.

%\textbf{We consider Herbertsmithite as a good candidate material for finding the spinon.
%Argue that it is a gapless spin liquid.}
%\textbf{ Spinons in kagome antiferromagnets.
% Hightlight Dirac state because it is gapless and that even though there is 
%good evidence that the ground state of the ideal kagome 
%antiferromagnet has a gap (Yan, Huse White) Herbertsmithe may be sufficiently far away.}

One exciting candidate material for a future ARPES experiment is
Herbertsmithite (ZnCu$_3($OH$)_6$Cl$_2$) which has got a lot of attention
in recent years. The absence of magnetic 
order\cite{Shores} down to 50 mK despite a Curie-Weiss temperature extracted 
from susceptibility measurements of around 300 K makes it one of 
the promising candidates for a quantum spin liquid in two dimensions. 
Herbertsmithite is a layered Mott insulator with the Copper atoms 
arranged in a Kagome lattice each carrying a spin 1/2 moment much 
like in undoped cuprates. As is well known the Kagome lattice is the 
most frustrated lattice having the least number of neighbours among 
two-dimensional lattices with triangular plaquettes which provide 
geometrical frustration. In multiple magnetic susceptibilty and 
heat capacity experiments\cite{Shores,Helton1,Vries,Helton2}, the strong magnetic field 
dependence of the linear-in-(low)temperature specific heat suggests 
that the heat carriers are spin-1/2 fermions thereby making a 
possible case for Herbertsmithite being a gapless spin liquid. 
Also, neutron scattering measurements on powder samples have 
found no evidence of a spin gap\cite{Helton1}. Perhaps the strongest evidence 
for spinons, however, comes from a recent neutron scattering study\cite{YSLeeTalk} on 
single crystals\cite{LeeSingleCrystal1,LeeSingleCrystal2} which found evidence 
for a spinon continuum but little evidence of a threshold for spinon 
production in momentum space. If ARPES could be applied to this 
material then it could provide complimentary evidence for spinons 
and solidify the case for their existence.

%\begin{figure}
%\includegraphics[width=0.7\linewidth]{FIGS/herbertsmithite1.eps}
%\caption{Herbertsmithite structure and Kagome layer chemistry}
%\label{fig:Kagome}
%\end{figure}

%explt refs
%[6] M.P. Shores et al., J. Am. Chem. Soc. 127, 13462 (2005)... original susceptibilty paper \theta_CW = -300 K
%[7] J.S. Helton et al., Phys. Rev. Lett. 98, 107204 (2007). ... susceptibility and sp. heat meas
%[8] M.A. de Vries et al., Phys. Rev. Lett. 100, 157205 (2008)... sp. heat meas
%[9] J.S. Helton et al., Phys. Rev. Lett. 104, 147201 (2010)....

There is also substantial evidence for spinons in Kagome antiferromagnets 
from theoretical calculations. Recently, Kagome antiferromagnets have 
been shown to have a $Z_2$ spin liquid ground state in large scale 
DMRG calculations with a gap to all excitations\cite{YanHuse}.  This 
strongly suggests that the excitations are spinons. However, these 
spinons may not be the spinons observed in the above experiments 
since they have not seen any evidence for a gap. Likely, this 
implies that Herbertsmithite is far 
enough away from the ideal nearest-neighbor Kagome Heisenberg AFM 
with its putative gapped ground state through the presence of 
perturbations such as those due to Dzyaloshinskii-Moriya (DM) 
interactions and impurities\cite{Shores,Helton1,Vries,Helton2}.  
Two natural candidates 
for a description of the phenomenology of Herbertsmithie are then 
a quantum critical point on the verge of magnetic ordering arising 
from the DM interactions
\cite{Cepas} 
and the gapless Dirac spin liquid(DSL) state\cite{Hastings, Ran} 
that is strongly supported by VMC studies\cite{Ran,Yasir}.

%theory refs
%bosonic spinons
%[6] S. Sachdev, Phys. Rev. B 45, 12 377 (1992). 
%[8] F. Wang and A. Vishwanath, cond-mat/0608129.
%[23] S. Isakov et al., Phys. Rev. B 72, 174417 (2005).
%fermionic spinons
%[24] S. Lee and P. Lee, Phys. Rev. Lett. 95, 036403 (2005).
%[25] O. I. Motrunich, Phys. Rev. B 72, 045105 (2005).
%Dirac Spinons
%[26] Y. Ran et al., cond-mat/0611414.

%M. B. Hastings ...Phys. Rev. B 63, 14413 (2000)... and intro for early refs

%\textbf{In this paper ... This is where we state the results. 
%Mention something about our discussion of fluctuations around mean field 
%theory is not complete (compactness vs non-compactness)... this is moved to
%organization paragraph.}

In this paper, we study the effect of Dirac spinons on the 
signatures of a Angle-Resolved Photoemission Spectroscopy(ARPES) 
on Herbertsmite. Even though the properties of Herbertsmithite's 
ground state are still unknown,  we will focus on the DSL 
or its generalization in the form of Algebraic Spin Liquids(ASL) 
both because it is a candidate description for Herbertsmithite and 
by itself is a paradigmatic state of matter with generic critical 
correlations. For example, the ASL state has also been considered 
in the context of underdoped cuprates\cite{RantnerWen}. Our main finding 
is that Dirac spinons lead to a linear energy dependence in the 
ARPES spectrum for low energies even though one would expect a 
sub-linear dependence from the anomalous exponent of the spinon propagator
that leads to the aforesaid critical correlations.

%\textbf{ Organization of paper: here we guide different readers through the 
%paper and help, for example, expermentalists choose which sections to focus on.}

To expose this experimentally falsifiable linear low-energy behaviour 
of the ARPES spectrum, the paper is organised as follows: In section I, 
we describe briefly the model Hamiltonian and its description in the 
slave boson picture. In Section II, we look at the mean-field theory 
of the Dirac state and calculate the ARPES spectral function for this 
state. This section may be of particular interest to experimentalists 
since it contains our key prediction regarding the ARPES signature 
of the Dirac spinons. In Section III, we go beyond mean-field and 
see how fluctuations of the mean-field do not modify the energy dependence 
of the mean-field ARPES spectral function. In our study, we have focused 
on small fluctuations of the phases of the mean-field and thus our 
study is incomplete with regard to large ($O[2\pi]$) phase fluctuations. 
Technically speaking, the gauge field that models these beyond 
mean-field phase fluctuations is non-compact in our calculation. 
We conclude in Section IV by discussing our results - especially 
pointing out the fact that our prediction might be unique 
to the Dirac state - and possible other experiments to pin 
down the putative spin-charge seperated excitations of the spin liquid.

While composing this manuscript, a recent similar study with 
similar conclusions has come to our 
attention\cite{Tang}. These authors also consider a Fermi surface state relevant 
for organic spin liquid candidate materials that is not discussed here.

\section{Model}
	Consider the $t$-$J$ model on the Kagome lattice. 
As is usual, the underlying model is the Hubbard model in the large-$U$ limit. 
The Hubbard model with a single site $U$ term is a good 
starting point for the spin physics of the Mott insulator Herbertsmithite 
similar to undoped cuprates. 
The $t$-$J$ Hamiltonian is
\begin{equation}\label{eq:tjmodel}
\Hamil = - t \sum_{\langle \vect{r},\vect{r}' \rangle, \sigma} c_{\vect{r},\sigma}^{\dagger} c_{\vect{r}',\sigma}
	 + J \sum_{\langle \vect{r},\vect{r}' \rangle} \spin_{\vect{r}} \cdot \spin_{\vect{r}'}
         -\mu_e \sum_{\vec r,\sigma} c^\dagger_{\vec r,\sigma}c_{\vec r, \sigma}	
\end{equation}
where the sum is over nearest-neighbour pairs of sites on the kagome lattice shown in Fig. \ref{fig:DSLBZ}, $\sigma$ is the spin index and $\spin = \sum_{\sigma,\sigma'} c_\sigma^\dagger \vec{\tau}_{\sigma,\sigma'} 
c_{\sigma'}$ with $ \vec{\tau}$ the vector of Pauli matrices. This model is understood within the constrained Hilbert space of one or less electron per site which can be written as
\begin{equation}\label{eq:tjconstraint}
\sum_{\sigma} c_{\vect{r},\sigma}^{\dagger} c_{\vect{r},\sigma} \leq 1 
\end{equation}
for any $\vect{r}$. To handle this constraint mathematically, the slave-boson approach \cite{Barnes,Coleman} is often used. This approach consists of increasing the Hilbert space and, in the process, converting the inequality
constraint to an equality constraint. Also, the slave-boson approach is naturally tailored to accomodate spin liquid ground state ansatzes[Reference: Baskaran, Zhou and Anderson]. 

The slave-boson approach starts with ``attaching" a boson to an fermion to compose an electron in the following way
\begin{equation}
c_{\vec r\sigma} = b^\dagger_{\vec r} f_{\vec r\sigma}
\end{equation}
where $b^\dagger_{\vec r}$ is called a ``holon" for it creates positive charge and $f_{\vec r\sigma}$ is called a ``spinon" and carry the spin index. Here, $b$ satisfies a bosonic commutation relations while $f$ satisfies a fermionic anti-commutation relations and their product satisfies the fermionic anti-commutation relations expected for the electron. Inserting these expressions into \eqref{eq:tjmodel}, the $t$-$J$ Hamiltonian in the slave-boson language becomes 
\begin{multline}
\Hamil = - t \sum_{\langle \vect{r},\vect{r}' \rangle, \sigma} b_{\vect{r}} f_{\vect{r},\sigma}^{\dagger} f_{\vect{r}',\sigma} b_{\vect{r}'}^\dagger 
         - 2J \sum_{\langle \vect{r},\vect{r}' \rangle} \spin_{\vect{r}} \cdot\spin_{\vect{r}'}\\
 -(\mu_e/2)\sum_{\vec r}\left(\sum_\sigma f^\dagger_{\vec r\sigma}f_{\vec r\sigma} -b^\dagger_{\vec r}b_{\vec r}+1\right)
\end{multline}
where  $\spin = \sum_{\sigma,\sigma'} f_\sigma^\dagger \vec{\sigma}_{\sigma,\sigma'}
f_{\sigma'}$. The equivalence of this representation 
of $\spin$ and the earlier one in terms of $c$ can be checked by comparing their matrix elements in the constrained Hilbert space defined by Eq. \eqref{eq:tjconstraint}. In addition to the new form of the Hamiltonian, the constraint now becomes an equality irrespective of the filling
\begin{equation}\label{eq:sbconstraint}
\sum_{\sigma} f_{\vect{r},\sigma}^{\dagger} f_{\vect{r},\sigma} + b_{\vect{r}}^\dagger b_{\vect{r}} = 1
\end{equation}
at each $\vect{r}$. It is now in a form where we can bring to bear the machinery of Lagrange multipliers.

\section{Mean-Field Theory}
\begin{figure}
\includegraphics[width=0.5\linewidth]{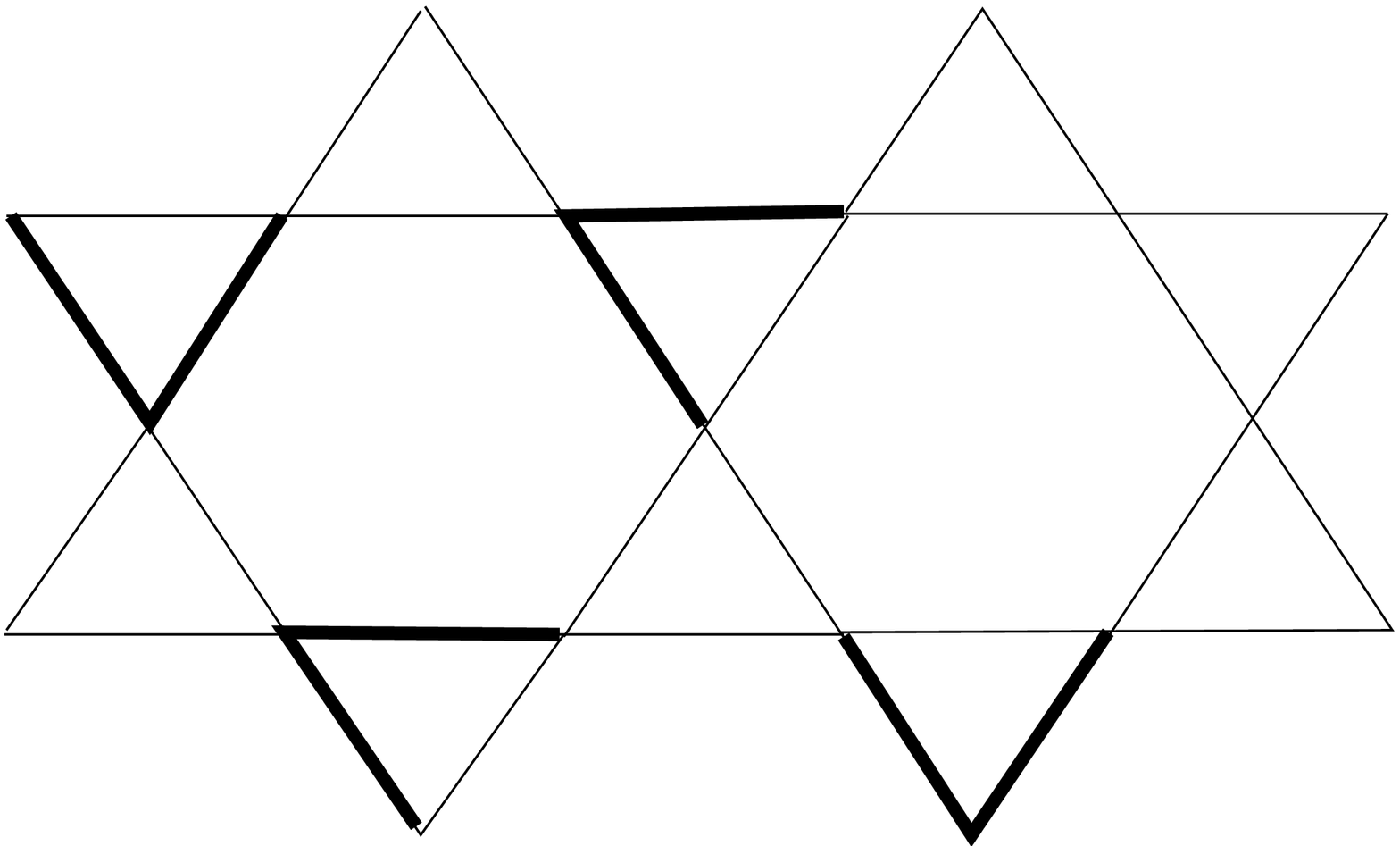}
\includegraphics[width=0.4\linewidth]{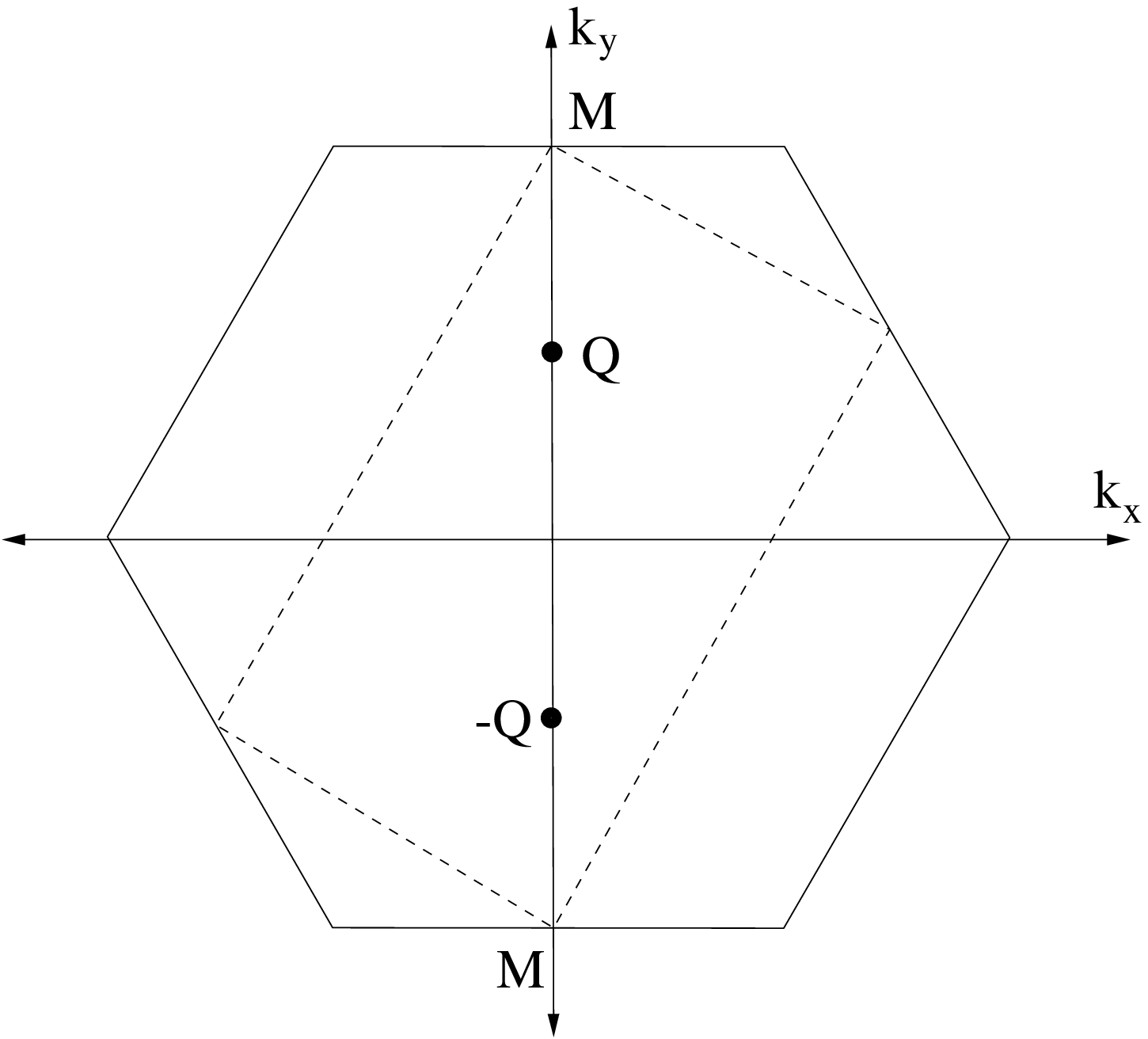}
\caption{In the left figure is illustrated the choice of gauge for the 
phases for the DSL mean field $s_{\vect{r}\vect{r}'}$;
thick lines correspond to a negative sign while thin lines correspond to a positive sign. This choice
(same as Ref. \onlinecite{Hermele2}) leads to a doubling of the unit-cell. Though the gauge-invariant
fluxes ($+1$ through the triangles, $-1$ through the hexagons) respect the underlying symmetry of the
kagome lattice as they ought to in a spin liquid. In the right figure, we show the Brillouin zones - 
the hexagon is the original Brillouin zone and the dashed rectangle is halved Brillouin zone due
to the doubling of the unit-cell. $Q$ and $-Q$ are the momenta at which the spinon chemical potential
lies at half-filling. The band structure leads to Dirac cones at these momenta as shown in
Fig. \ref{fig:DSLband}.}
\label{fig:DSLBZ}
\end{figure}

As mentioned in the Introduction, we choose Dirac Spin Liquid as our mean field ansatz. The quartic terms of
the Hamiltonian are decoupled using an auxiliary field and written as follows :
\begin{multline}
\Hamil = - t \chi_0\sum_{\langle \vect{r},\vect{r}' \rangle, \sigma} 
   s_{\vect{r}\vect{r}'} b_{\vect{r}}^\dagger b_{\vect{r}'}
         - J \chi_0\sum_{\langle \vect{r},\vect{r}' \rangle,\sigma} 
s_{\vect{r}\vect{r}'} f_{\vect{r},\sigma}^\dagger f_{\vect{r}',\sigma} \\
  -\sum_{\vec r} \left(\mu_f\sum_{\sigma} f_{\vect{r},\sigma}^{\dagger} f_{\vect{r},\sigma} +\mu_b b_{\vect{r}}^\dagger b_{\vect{r}}\right)%+\frac{\mu_f-\mu_b}{2}
\label{eq:meanfieldham}
\end{multline}
where $\chi_0$, $\mu_f$ and $\mu_b$ are mean field parameters and $s_{\vec r,\vec r'}=\pm1$ according to the phases defined in Fig. \ref{fig:DSLBZ}. The self-consistency condition of the mean-field theory is then given by 
\begin{eqnarray}
   |\langle f_{\vect{r},\sigma}^\dagger f_{\vect{r}',\sigma}\rangle| &=& \chi_0 \text{  for n.n. }\vect{r},\vect{r}'\\
   \langle  f_{\vect{r},\sigma}^\dagger f_{\vect{r},\sigma}\rangle &=& 1-x\\
   \langle  b_{\vect{r},\sigma}^\dagger b_{\vect{r},\sigma}\rangle &=& x
\end{eqnarray}
where the expectation is evaluated in the ground state and $\mu_b$ and $\mu_f$ are chosen 
to impose the latter two constraints and replace $\mu$ and the constraint of 
Eq. \eqref{eq:sbconstraint} . The DSL mean-field ansatz corresponds
effectively to a background magnetic flux threaded through the Kagome lattice 
as shown in the Fig. \ref{fig:DSLBZ}. It is chosen so that
$\pi$-flux pierces the hexagons and zero flux through 
the triangular plaquettes. As we see in Fig. \ref{fig:DSLBZ}, the flux still respects
the symmetry of the lattice even though the particular choice of gauge 
or phases for this ansatz has a doubled unit cell.

%Since we aim to calculate the ARPES spectral function for a half-filled Kagome AFM, we should also state the self-consistency relating to the half-filling constraint. $\langle  f_{\vect{r},\sigma}^\dagger f_{\vect{r},\sigma} \rangle = 1$. At mean-field the half-filling constraint is only satisfied on an average because the fluctuation of the spinon occupation number for this ansatz is non-zero. And including the fluctuations around the mean field better take care of the constraint exactly by making the fluctuation in the spinon occupation and all its higher moments to zero. Also, another self-consistency condition for this ansatz is $\langle \spin_{\vect{r}} \rangle = 0$. This is the spin liquid condition.

%The Kagome lattice has a three-site unit cell and following Ref. [\onlinecite{Ran}] (cf. Fig. 1 (a))  - we label the sites by pairs $(\vect{R},i)$, where $\vect{R} = n_1 \vect{a}_1 + n_2 \vect{a}_2$ is the lattice vector labeling the unit cell, and $i = 0,1,2$ labels the three site Fig. \ref{fig:Kagome}. We choose $\vect{a}_1 = \unit{x}$ and $\vect{a}_2 = (1/2)\unit{x}  + (\sqrt{3}/2) \unit{y}$, so the distance between nearest-neighbour sites is $1/2$. The reciprocal lattice primitive vectors can be chosen as $\vect{b}_1 = 2 \pi [\unit{x} - (1/\sqrt{3}) \unit{y}]$ and $\vect{b}_2 = (4 \pi/\sqrt{3}) \unit{y}$, and the Brillouin zone is as shown in Fig. \ref{fig:DSLBZ}. 

\begin{figure}
\includegraphics[width=0.48\linewidth]{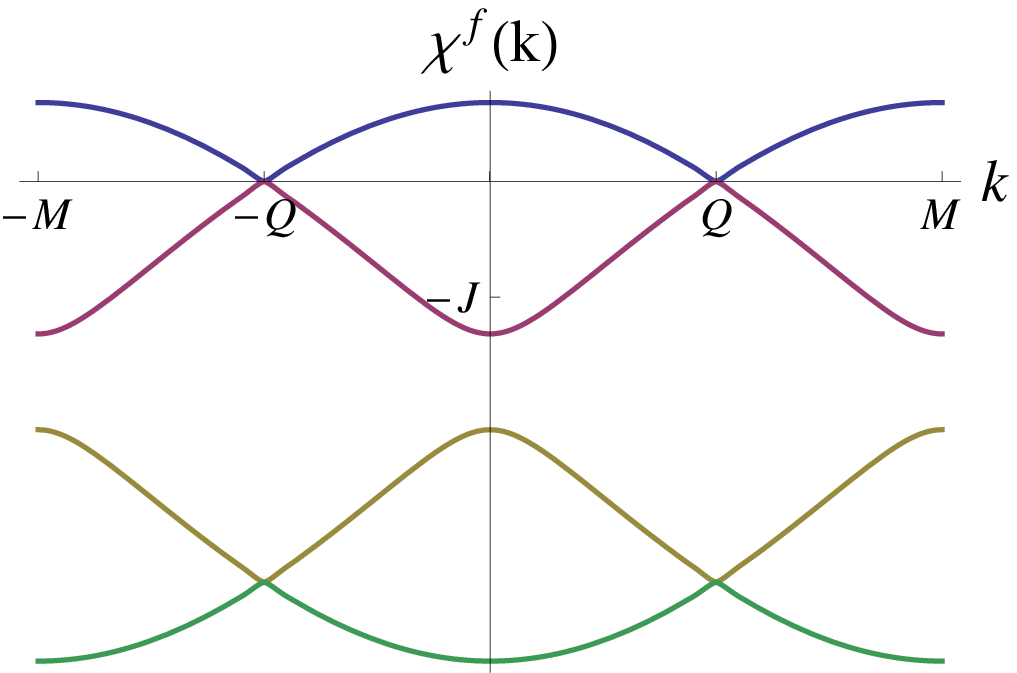}
\includegraphics[width=0.5\linewidth]{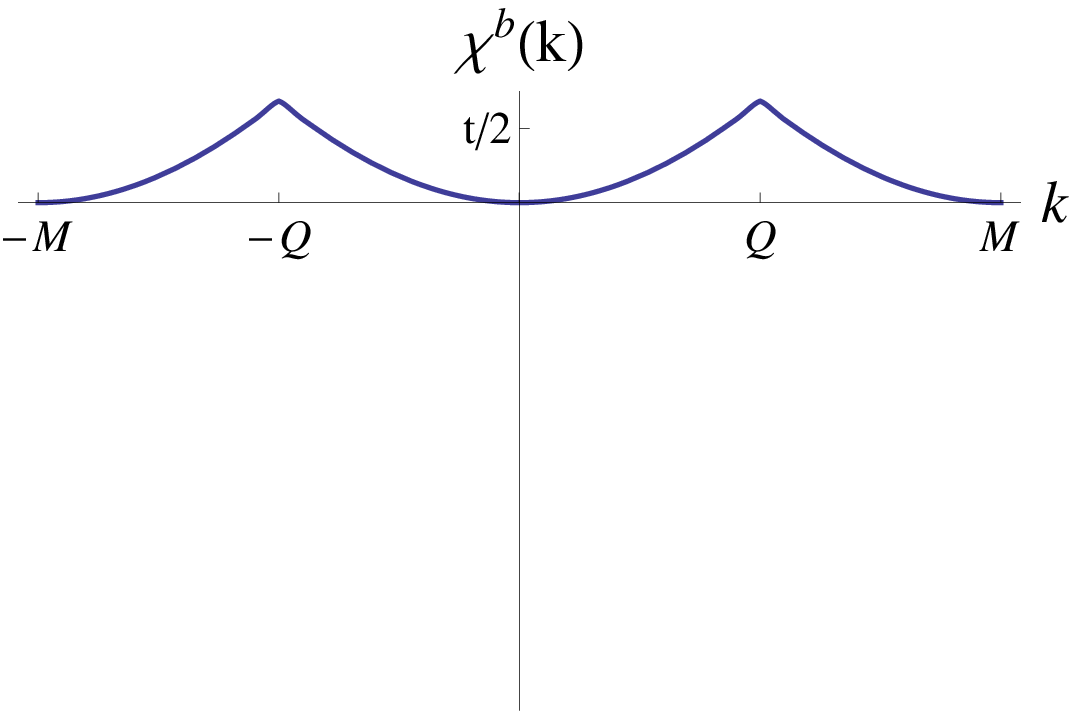}
\caption{In this figure, we show the mean field band structure for the Spinons
and Holons. On the left we show the lowest four bands (there are two flat bands at higher energy
which are not shown in the figure). Of these four bands, at half-filling, the lowest
three fill out thus leading to Dirac cones for the dispersion of the low energy spinons.
On the right we show the lowest band for the holon which is unoccupied as there are
no holons at half-filling.}
\label{fig:DSLband}
\end{figure}

Let us now turn to the spectrum of this Hamiltonian. We notice that the 
resulting quadratic Hamiltonian of Eq. \ref{eq:meanfieldham} consists of 
non-interacting holons and non-interaction spinons hopping in the presence 
of the same background magnetic flux. The only difference that is the 
energy scale associated with holons is $t$ while for spinons it is $J$. 
With this in mind, we diagonalize the Hamiltonian following the notation 
of Fig. 1(a) in Ref. [\onlinecite{Ran}] where Bravais lattice vectors 
are chosen to be $\vect{a}_1 = \unit{x}$ and $\vect{a}_2 = (1/2)\unit{x} + (\sqrt{3}/2) \unit{y}$ 
giving the Brilliouin zone shown in Fig. \ref{fig:DSLBZ}. When this quadratic Hamiltonian 
is solved we get the band structure shown in Fig. \ref{fig:DSLband} with Dirac-like nodes as the fermi surface.

Finally, the question of the holon chemical potential $\mu_b$ needs some attention. 
We assume that the holon chemical potential is chosen so that the doping $x=0$ and the bosons remain uncondensed at zero temperature. These assumptions appear quite valid for an ARPES experiment on Herbertsmithite so long as the density $x$ of induced holes created by photoemission remains small. It turns out, to achieve this at zero temperature, any negative value of $\mu_b$ will suffice. However, a specific value will be chosen at finite temperatures in the experimentally relevant regime. We will therefore leave $\mu_b$ as a finite negative valued parameter in our theory. This produces the spectrum for the holons on the right side of Fig. \ref{fig:DSLband}.

%What choices are afforded to us for the holon chemical potential that keeps the holon bands empty? At zero temperature, any value below the bottom of the lowest band would suffice. In order to make it lowest energy wavefunction upon introduction of  the holon(which has to be accompanied by the removal of a spinon) we would like to match the bottom of the holon bands with spinon chemical potential. Since if the bottom of the holon band were to remain at an energy value above the spinon chemical potential, we can always tune the holon chemical potential to bring the holon band down in the process affording us a new wavefunction(with one holon and half-filling minus one spinons) which has a lower energy. 

\subsection{The Spectral Function}
Let us now turn to the mean field predictions for the ARPES spectral function defined by 
\begin{multline}
A(\vect{k},\omega) = \\
   -\frac{1}{\pi} \text{Im}\left[\int d\vec{r} \int_{-\infty}^0 dt \left\langle 
   c_{\vect{r},\sigma}^\dagger(t) c_{\vect{0}}(0) \right\rangle
   e^{- i \dotp{r}{k} - i \omega t} \right].
\label{eq:Spec}
\end{multline}
At the mean field level, this becomes 
\begin{equation}
A_{mean}(\vect{k};\omega) \propto \sum_{\lambda_1,\lambda_2} 
\int dq\, \delta\left(\omega - 2t \en^b_{\lambda_1}\left(\vect{k}-\vect{q}\right) + J \en^f_{\lambda_2}\left(\vect{q}\right)\right)
\label{eq:Specfn}
\end{equation}
where $\en^b$,$\en^f$ are the holon and spinon bands respectively and
$\lambda_1$, $\lambda_2$ are band indices. The above expression has a
 clear interpretation: all the possible spinon-holon
combinations with the right energy-momentum values combine to 
give the requisite hole at wavevector $\vect{k}$ and 
energy $\omega$ in the ARPES experiment. The factor of two with the 
holon or $t$ term in the above equation comes from spin
counting but we need not overly worry about such factors. Ultimately, we are 
concerned with the $\omega$ or $\vect{k}$ dependence
of the spectral function, and the functional form of this 
dependence is not going to be affected by such factors of two.

\begin{figure}
\includegraphics[width=0.7\linewidth]{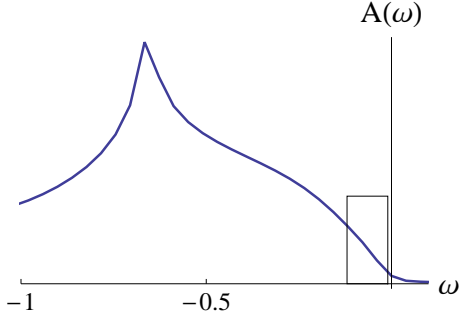}
\caption{In this figure, we show a numerically calculated 
representative spectrum for ARPES at the mean-field level. The numerical
calculation confirms the linear behavior
at low energies (highlighted using the rectangle) as argued for in the text (Sec III). 
Our choice of parameters were $J=0.5$ and $t=1$.}
\label{fig:Akmean}
\end{figure}

We are interested in the low energy behavior of the system in general and the 
spectral function in particular. We can easily
see for the mean-field bands, most of the bands won't 
play any role in the low energy behavior. It is only the highest
filled spinon band and the lowest empty holon band that will 
contribute to the $\omega \rightarrow 0$ limit. For these bands,
the low $\omega$ contribution will come from near specific 
wavevectors that connects the ``Dirac" points of spinon bands
to the minima of the holon bands. Near these specific wavevectors, 
the spinon and holon spectrum are linear 
and quadratic in $|\vect{q}|$. Thus, for smaller and 
smaller $\omega$, the linear term starts dominating in the integral
and we get $A_{mean} \propto \omega$ for $\omega \rightarrow 0$. We show 
in Fig. \ref{fig:Akmean} a numerical confirmation of the 
mean-field result. In addition, a back-of-envelope estimate tell us
that linearity is expected 
up to about $\frac{t^2}{4J}\frac{\chi'(Q_f)^2}{\chi'(Q_b)}$
where $Q_f$ and $Q_b$ are the momenta at which the Dirac point
and the holon minimum sit respectively. 
This linear behavior in energy is---if the mean-field theory 
were qualitatively correct---a key signature easily extracted from ARPES data. 

\subsection{Continuum limit and scaling analysis}
To understand fluctuations around mean field theory, 
a continuum limit approximation is essential. To this end, it is 
useful to carry this out first directly at the mean field level. Such an analysis will thus
facilitate the scaling discussion when we go beyond mean-field theory. We define
the continuum fields through
%Update this section with a discussion of the continuum limits:
\begin{eqnarray}
 b_{\vec{R},d} & = & \sum_n \phi_n(\vec{R}) U^n_{d} e^{i Q_b^n.(\vec{R})} \nonumber \\
 f_{\vec{R},d,\sigma} & = & \sum_{m,\lambda} \psi_{m,\lambda,\sigma}(\vec{R})
 U^m_{\lambda,d}e^{iQ_f^m.(\vec{R})}
\label{eq:ContinuumFields}
\end{eqnarray}
where $\vec{R}$ refers to unit-cell index and $d$ refers to the index of
the bases within the unit-cell.
The sum is understood to be restricted to the low-energy modes,
$n$ refers to the index characterizing the (three) holon minima in the holon band structure
and $Q_b^n$ are the momenta in the Brillouin zone where the minima sit respectively, 
$m$ refers to the index characterizing the (two) Dirac points in the spinon band structure
and $Q_f^n$ are the momenta in the Brillouin zone where the Dirac points sit respectively,
$\sigma$ is the spin index, and $\lambda$ is the ``Dirac" index characterizing
the branches of the Dirac cone. $U^n_{d}$ and $U^m_{\lambda,d}$ are the eigenmodes
of the bandstructure corresponding to holon minima and Dirac points respectively.

Thus, using the definition of Eq. \ref{eq:Spec}  we can get for the continuum spectral function
\begin{equation}\label{eq:ConSpec} 
A(Q_f^m-Q_b^n+q,\omega) = \sum_{\lambda,\sigma}M^{mn}_{\lambda,\sigma} A^{m,n}_{\lambda,\sigma}(q,\omega)
\end{equation}
where $M^{mn}_{\lambda,\sigma}$ are contributions due to the eigen modes
$U^n_{d}$ and $U^m_{\lambda,\vec{d}}$. $M^{mn}_{\lambda,\sigma}$ is independent of
momentum and energy, and $A^{m,n}_{\lambda,\sigma}(q,\omega)$ contains all
the momentum and energy dependence which is what we will focus on in the following.

With the definitions in Eq. \ref{eq:ContinuumFields}, the low energy physics (of the
quadratized Hamiltonian Eq. \ref{eq:meanfieldham}) is described by 
an effective 2+1 continuum field theory where the 
spinons are described by a Dirac fermionic field (due to the linear spectrum 
near the spinon chemical potential) $\Psi$ - a \emph{linear} combination
of $\psi_{\lambda}$, details of which can be found in Appendix A of
Ref. \onlinecite{Hermele2} - and the holons are described by a 
non-relativistic complex scalar field $\phi$ 
(due to the quadratic spectrum near the holon chemical potential). 
%A derivation of the fermionic part can be found in Ref. [\onlinecite{Hermele}]. We sketch 
%the bosonic part in the appendix. 
The Lagrangian for this field theory is
\begin{multline}
L_{free}  =  \int dr dt \bigg[\bar{\Psi}(r,t) \gamma^\mu \partial_\mu \Psi(r,t) + \\\phi^*(r,t)
 (i \partial_t + \partial_r^2 / 2 m_{holon} - \mu_b) \phi(r,t) \bigg]
\label{eq:MFFT}
\end{multline}
The mean field fixed point is then determined by how these fields scale when space-time is scaled (and from this field scalings, we can write a scaling 
form for the Spectral function). When we scale as
$r \rightarrow b r ; t \rightarrow b t$,  it can be shown that scaling the 
fields as $\Psi \rightarrow b^{-1} \Psi ; \phi \rightarrow b^{-1} \phi$ leaves the 
Lagrangian invariant. Using these field and space-time scalings and the continuum limit of Eq. \ref{eq:Spec}
\begin{multline}
A_{mean}(\vect{k},\omega) \propto \\
\text{Im}\int dr \int_{-\infty}^0 dt \langle \phi(\vect{r},t) \bar{\Psi}(\vect{r},t) 
\Psi(\vect{0},0) \phi^{*}(\vect{0},0) \rangle
 e^{- i \dotp{r}{k} - i \omega t},
\label{eq:Specfield}
\end{multline}
we can show that spectral function scales as $ A_{mean}(b^{-1} \omega, b^{-1} k) = b^{-1} A_{mean}(\omega, k)$ which is consistent with $A_{mean} \propto \omega$ as we obtained from more elementary considerations.

\section{Beyond Mean-field Theory}

%\subsection{The U(1) Gauge structure}
The $t$-$J$ Hamiltonian remains invariant under a local gauge transformation made to the spinons and holons
\begin{equation}
    b_i \rightarrow e^{i \theta_i} b_i,\quad f_{i,\sigma} \rightarrow e^{i \theta_i} f_{i,\sigma}.
\end{equation} 
Thus, when we include the fluctuations beyond the mean field level, the resulting continuum Lagrangian should remain invariant under this transformation. As is well know, this is achieved through replacing the derivative by a covariant derivative $\partial_\mu \rightarrow D_\mu = \partial_\mu - i e A_\mu$ where $e$ is the electric charge and $A_\mu$ is a Maxwell vector potential satisfying the usual transformation law: $A_\mu(\vect{r},t) \rightarrow A_\mu(\vect{r},t)
 - \partial_\mu \theta(\vect{r},t)$ when $\Psi(\vect{r},t) \rightarrow e^{i \theta(\vect{r},t)} \Psi(\vect{r},t)$
and $\phi(\vect{r},t) \rightarrow e^{i \theta(\vect{r},t)} \phi(\vect{r},t)$. Thus, the beyond mean-field Lagrangian for the spinons and holons is
\begin{multline}
L  =  \int dr dt \bigg[\bar{\Psi}(r,t) \gamma^\mu D_\mu \Psi(r,t) -\frac{1}{4} F_{\mu\nu}F_{\mu\nu}  \\
 \phi^*(r,t)(i D_t + (D_x^2 + D_y^2) / 2 m - \mu) \phi(r,t) \bigg]
\label{eq:beyondMFFT}
\end{multline}
where $F_{\mu\nu} = \partial_\mu A_\nu - \partial_\nu A_\mu$. 

%By taking a large-N limit, discussed in the next subsection, we will using this Lagrangian to study both the stability of the Dirac fixed point discovered in Ref. \onlinecite{Hermele2008}??????, the holon propagator in the following section and finally turn to the central result of this paper, that the electron propagator is not strongly renormalized from its mean field form by the gauge fluctuations.

%The photon propagator is worked out in standard references\cite{FieldTheory}.

%\subsection{The Large-N limit}
%We start by summarzing the beyond mean-field result of the spinons. 
Now, Dirac fermions coupled to an $U(1)$ gauge field in $2+1$ dimensions is a strongly-coupled problem. 
To make progress, the gauge field contributions are typically handled by a large-N perturbation 
series \cite{Appelquist,KimLee,RantnerWen, Hermele0, Hermele1, Franz, Kaveh}
arising from $N$ flavors of the Dirac fermions. In this case, 
for large $N$, $1/N$ provides a formal small parameter provided the 
coupling scales with $N$ as $e^2 \sim 1/N$. When the pertubation theory 
is carried out, one finds that the spinon propagator acquires an anomalous dimension because of the gauge field coupling $e$ (See e.g. Ref. [\onlinecite{Hermele1}]'s Appendix B). Formally, when we scale space-time as
$r \rightarrow b r ; t \rightarrow b t$,  the Dirac field scales as $\Psi \rightarrow b^{-1-\delta} \Psi$ where $\delta$ is the anomalous dimension at the fixed point. This anomalous dimension is the manifestation of the fact that the gauge-field contributions make the theory flow to a new fixed point (with different scaling) away from the mean-field fixed point. The general intuition is that the strength of fluctuations will only increase when $N$ is decreased and thus this new fixed point (called Algebraic fixed point) may dominate a large portion of the low energy sector for the physical case of $N=4$.

%We want to go beyond mean-field theory to understand what role do the fluctuations of the mean-field play in the process
%of exactly enforcing the single occupancy constraint. The low energy fluctuations are the fluctuations in the phases
%of the mean-field solution. These fluctuations can be taken into account exactly by incorporating an $U(1)$ gauge field which
%couples to the gradient operator leading to $\partial t \rightarrow \partial t - A_0(t)$ and $\partial i \rightarrow 
%\partial t - A_i(t)$. 
Our aim is to understand the effect of the beyond mean-field fluctuations on the holons, a subject not visited in previous studies due mostly to a focus on the Heisenberg model. In an earlier work\cite{RantnerWen}, the electron spectral function for the Dirac Spin Liquid was considered in the context of underdoped cuprates, but this study differs significantly from ours in that they assumed that the bosons were condensed as they were motivated by pseudogap phenomena at finite doping.
%We notice that since the bosonic lagrangian is
%linear the $\partial t$, the $A_0(t)$ field just leads to a shift in the bosonic chemical potential. 
To reiterate our main assumption, we will suppose that the bosons are \emph{uncondensed}, i.e. $\mu < 0$, an assumption that seems reasonable if a successful ARPES experiment is to be performed.  In the ideal case at half-filling, under the exact single occupancy constraint, there are no bosons around. When an ARPES photon ejects an electron, a boson is created before it gets destroyed by the replenishing of the electron through an electronic bath connected to the sample. This  boson is then uncondensed at the temperature of the experiment because they are  very dilute. If the electrons fail to be replenished, however, the bosons density could grow to a level where they condense. The resulting charged surface, though, would then affect the trajectory of the ejected electron and the ARPES experiment would fail to obtain meaningful data. Hence, a successful ARPES experiment would probe the undoped insulator only when the bosons remain uncondensed.
%We are assuming that the shifted bosonic chemical potential keeps the bottom
%of the holon band matched to the spinon chemical potential.

In the following Section IV. A, we discuss in detail the effect of gauge field on the bosonic field.
%In earlier works, (I'm thinking of Wen's paper and probably some of Sachdev's too .. will have to look up
%the later) there was work done along these lines but the bosons were assumed to be condensed because
%they were in the context of Hubbard model at finite doping, and the motivation was that the condensed
%boson would give rise to superconductivity(QUESTION and the uncondensed limit would to connect
%to the Mott state at low doping ?). 
We find that the gauge-field renormalized bosonic propagator does not acquire an anomalous dimension unlike the spinon propagator and still scales as a free boson, $\phi \rightarrow b^{-1} \phi$ upon $r \rightarrow b r ; t \rightarrow b t$. This is expected due to the finite energy cost of the bosonic excitations when they remain uncondensed.
% the mistaken conclusion prior to Jan '12
%The implication of this beyond mean-field result is that, at large-N, the mean-field prediction
%of $ A(b^{-1} \omega, b^{-1} k) = b^{-1} A(\omega, k)$ will be modified to $ A(b^{-1} \omega, b^{-1} k) 
%= b^{-1-\delta} A(\omega, k)$ or $A \propto \omega^{1+\delta}$. In previous works \cite{RantnerWen,
%Hermele1,Franz}, it was shown
%that $\delta = -4/(3 \pi^2 N)$. Thus, the beyond mean-field ARPES prediction for the Kagome
%Antiferromagnet would be a sub-linear power law in $\omega$ if the Dirac Spin Liquid is
%the ground state (See Fig. \ref{fig:Akbeyondmft} for an illustration).
Furthermore in Section IV B, we show that the electron propagator made of
the spinon and boson also retains its mean-field scaling upon
taking lowest order gauge-field corrections. This is intriguing when
contrasted with the anomalous scaling of the spinon.

%\textbf{MEMO : HIDE this ?}
%A Wilsonanian RG kind of hand-waving explanation would be as follows : at the Algebraic fixed point, for the Dirac fermion lagrangian to stay scale invariant, a thin shell of Energy-momentum has to be integrated out from $b \Lambda$ to $\Lambda$ and from $b \Lambda$ to $\Lambda$ and  the Dirac field has to rescale as $\Psi \rightarrow \Psi/b^{1+\delta}$(I've to make sure of this). Under this RG operation for the bosons, we see that the momentum shell integration makes the momentum shell thickness an order lower compared to the Energy shell and thus the $k^2$ term becomes irrelevant. The $E$ shell integration can just be absorbed in to a shift in chemical potential and the field doesn't have to rescale(same as for Gaussian fixed point in O(n) theory). Thus, the bosons remain as free bosons. In the appendix, we show that the self-energy correction to the bosonic propagtor can not have any logarithmic terms, thereby giving no anomalous exponent to the propagator.

%\begin{figure}
%\includegraphics[width=0.6\linewidth]{FIGS/Cartoon.eps}
%\caption{A cartoon of the spectral function beyond mean field}
%\label{fig:Akbeyondmft}
%\end{figure}

\subsection{Renormalized Perturbation Theory for Holons}
Let us now focus on the holon part of the Lagrangian---a non-relativistic complex bosonic field---that couples only to the $U(1)$ gauge field:
\begin{equation}
\mathscr{L}_\phi = \int d^2 r dt \phi^*_0(\vect{r},t)(i D_t + \frac{D^2_x+D^2_y}{2m_b} - \mu_b)\phi(\vect{r},t)
\label{eq:bareL}
\end{equation}
where $D_\mu = \partial_\mu - i eA_\mu(\vect{r},t)/sqrt{N}$ as stated earlier though now extended to arbitrary $N$
%\cite{e2footnote}
, the bare mass parameter $m_b$ characterises the curvature of the quadratic dispersion at the bottom of the holon band and $\mu_b$ is the bare holon chemical potential. With the assumption of uncondensed holons, $\mu_b$ is strictly negative, i.e. less than the bottom of holon band.

To understand the holon propagator in the large-N limit, it turns out we need to first understand the photon propagator. 
The bare photon propagator receives a finite renormalization due to both spinon bubbles and holon bubbles even at 
leading order in $1/N$.  However, the holon bubble does not affect the renormalization of the most singular 
contribution arising from the fermionic spinon vacuum bubble \cite{KimLee}. Thus, the O(1/N) corrected photon 
propagator is (see  Eqn B5 of Ref. [\onlinecite{Hermele1}])
\begin{equation}
\includegraphics[width=0.2\linewidth]{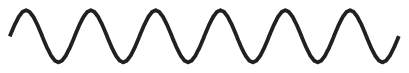}
= \frac{16}{N|q|}(\delta_{\mu\nu} + (\xi - 1)\frac{q_\mu q_\nu}{q^2})+\ldots
\label{eq:photon_propagator}
\end{equation}
where the ellipsis represents subdominant corrections that are smaller than the term shown as $q\to0$ and in 
the following we will use the Lorentz's gauge, $\xi = 1$. This Coulomb-like form is 
then the natural interaction that the holons feel.

%\begin{comment}
%\begin{eqnarray}
%\mathscr{L}_{free} & = &  \int d^2 r dt \phi^*_0(\vect{r},t)(i \partial_t + \frac{\partial^2_x+\partial^2_y}{2m_b}+\mu_b)
%\phi_o(\vect{r},t) \nonumber \\
%\mathscr{L}_{int} & = &  \int d^2 r dt \phi^*_0(\vect{r},t)(A_t(\vect{r},t) - i\frac{A_x(\vect{r},t) +A_y(\vect{r},t)}{m_b})
%\phi_o(\vect{r},t) \nonumber \\
%& & + \phi^*_0(\vect{r},t)( -\frac{A^2_x(\vect{r},t) +A^2_y(\vect{r},t)}{2m_b})\phi_o(\vect{r},t)
%\end{eqnarray}
%\end{comment}

For the perturbation theory we follow standard texts \cite{FieldTheory} and define the original (bare) Lagrangian with bare parameters in terms of the physical parameters (to be determined by the experiment). We start by defining a renormalized bosonic field in terms of the bare field
\begin{equation}
\phi_0 \equiv Z^{1/2} \phi \cdots Z \equiv \text{field rescaling of bosons}
\label{eq:fieldrescaling}
\end{equation}
Putting this in to Eqn. \ref{eq:bareL}, the Lagrangian looks like
\begin{multline}
\mathscr{L}_\phi  =  \int d^2 r dt \bigg[\phi^*(\vect{r},t)(i D_t + \frac{D^2_x+D^2_y}{2m} - \mu)\phi(\vect{r},t) \\
  + \phi^*(\vect{r},t) \frac{D^2_x+D^2_y}{2 \delta_m})\phi(\vect{r},t) \\
  + \delta_Z \phi^*(\vect{r},t) i D_t \phi(\vect{r},t) 
  + \delta_\mu  \phi^*(\vect{r},t) \phi(\vect{r},t)\bigg]
\label{eq:renormL}
\end{multline}
where $\delta_Z  =  Z - 1$, $\delta_m  = \left( \frac{1}{m}- \frac{Z}{m_b}\right)^{-1}$ and $\delta_\mu =  - Z\mu_b + \mu$ with $\mu$ a renormalized version of $\mu_b$ not to be confused with the chemical potential of the electrons. These are the counterterms. Since the holons do not renormalize the photon field, it's propagator
stays the same as in Eqn. \ref{eq:photon_propagator}.

We rewrite the renormalized Lagrangian as a free part
\begin{equation}
\mathscr{L}^r_{free} =  \int d^2 r dt \phi^*(\vect{r},t)(i \partial_t + \frac{\partial^2_x+\partial^2_y}{2m} - \mu)
\phi(\vect{r},t) 
\end{equation}
a counter term part 
\begin{multline}
 \mathscr{L}_{counter} =  \int d^2 r dt \bigg[\delta_z  \phi^*(\vect{r},t) i \partial_t \phi(\vect{r},t) + \\
 \phi^*(\vect{r},t)  \frac{\partial^2_x+\partial^2_y}{2\delta_m} \phi(\vect{r},t)  
+ \delta_\mu  \phi^*(\vect{r},t) \phi(\vect{r},t) \bigg]
\end{multline}
and an interaction part
\begin{multline}
 \mathscr{L}_{int} = \int d^2 r dt \bigg[ Z \phi^*(\vect{r},t)A_t(\vect{r},t)\phi(\vect{r},t)\\
 - iZ\phi^*(\vect{r},t) \left(\frac{\partial_x A_x(\vect{r},t) + 
      \partial_y A_y(\vect{r},t)}{m_b}\right) \phi(\vect{r},t) \\
 + \phi^*(\vect{r},t)\left( -Z\frac{A^2_x(\vect{r},t) +A^2_y(\vect{r},t)}{2m_b}\right)\phi(\vect{r},t)\bigg]
\end{multline}
Therefore, the Feynman rules are given by: the boson propagator line
\begin{equation}\label{eq:bpropagator}
\includegraphics[width=0.2\linewidth]{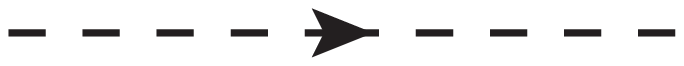}
= \frac{1}{\omega - \big(|\vect{k}|^2/2m + |\mu| \big) + i \epsilon}
\end{equation}
the counter terms by
\begin{eqnarray}
&& \includegraphics[width=0.2\linewidth]{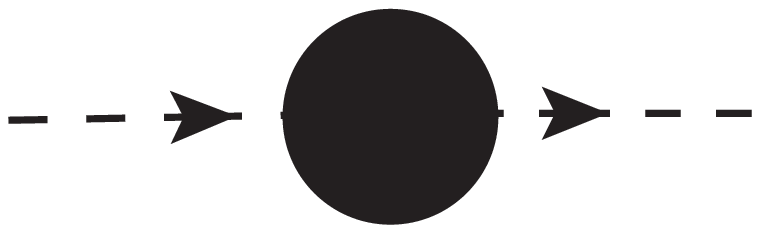}
= \delta_Z \omega  \nonumber\\
&& \includegraphics[width=0.2\linewidth]{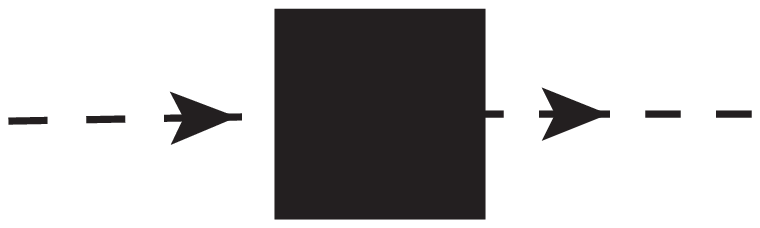}
= |\vect{k}|^2/2\delta_m  \nonumber\\
&& \includegraphics[width=0.2\linewidth]{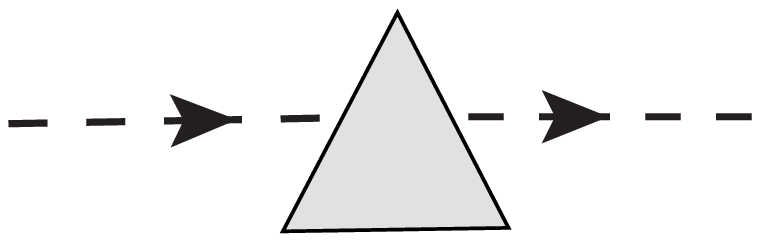}
= \delta_\mu
\end{eqnarray}
and the interaction vertices by
\begin{itemize}
\item Boson-Gauge Field's ``Time" component vertex 
\begin{equation}\label{eq:bosontimevertex}
 \includegraphics[width=0.2\linewidth]{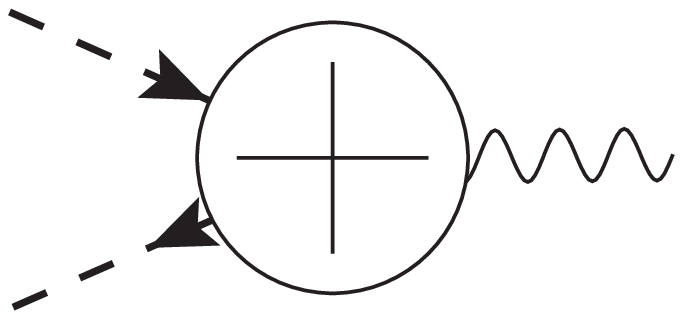}
= Z 
\end{equation}
\item Boson-Gauge Field's ``Space" component vertices 
\begin{equation}\label{eq:bosonspacevertex}
 \includegraphics[width=0.2\linewidth]{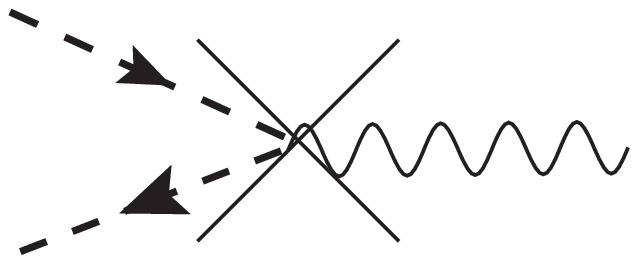}
 = Z k_i/m_b
\end{equation}
where $i$ refers to the Gauge field's ``space" component we are looking, i.e. if $\phi$
interacts with $A_x$, then the vertex will contribute $k_x$, etc. The are actually two such vertices of this type, one where $k_i$ goes with the incoming boson's ($\phi$) momentum and one where $k_i$ goes the outgoing boson's ($\phi^{*}$) momentum. We can collected all these vertices together by defining $V^{\mu}(\vec k)$ to be Eq. \eqref{eq:bosontimevertex} for $\mu=t$ and Eq. \eqref{eq:bosonspacevertex} for $\mu=x$ or $\mu=y$.
\item Boson-Gauge Field's quartic vertex 
\begin{equation}
\includegraphics[width=0.2\linewidth]{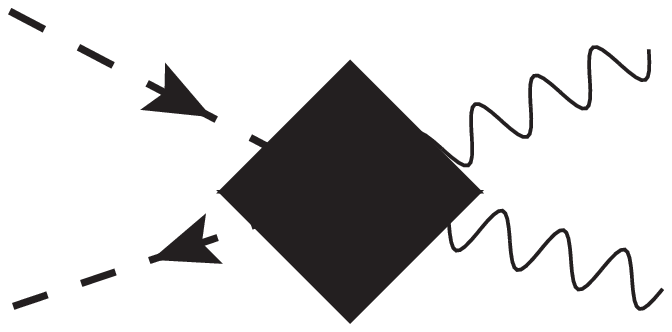}
= -Z/2m_b 
\end{equation}
\end{itemize}
Note: energy-momentum is conserved at the interaction vertices.

Now, we look at the 2-point free vertex function 
$\Gamma_0^2(\vect{k},\omega)=\omega -  \big(|\vect{k}|^2/2m + |\mu| \big)$ and impose our renormalization conditions on the renormalized 2-point function
\begin{eqnarray}
\Gamma^2(\vect{k}=(0,0),\omega=M)& = & \mu=-|\mu| \nonumber\\
d\Gamma^2/d\omega(\vect{k}=(0,0),\omega=M) & = & 1 \nonumber\\
d\Gamma^2/dk^2(\vect{k}=(0,0),\omega=M) & = & -1/2m
\label{eq:RG_cond}
\end{eqnarray}
In other words, we chose the time-like scale $(\vect{k},\omega)=((0,0),M)$ as the renormalization scale. 
% This scale is representative of the infra-red limit at the Dirac point in the Brillouin zone.
% because
%1) for massless QED(which is the case for the Dirac spinons), that is a good prescription
%\cite{FieldTheory}, and 2) with the same prescription,
%the bosonic integrals are regularised as well.

\begin{figure}
\includegraphics[width=0.9\linewidth]{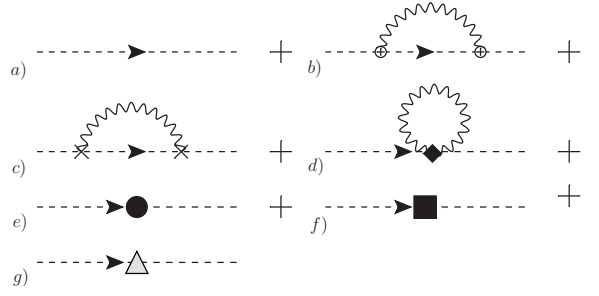}
\caption{Perturbation Series for Bosons to first order in $O[1/N]$}
\label{fig:Pertdiag}
\end{figure}

With the above setup, let us now turn to our main interest, the holon's anomalous dimension which is governed by the holon field rescaling $Z$
\cite{FieldTheory}. We will consider this to order $1/N$ through the diagrams of the propagator shown in Fig. \ref{fig:Pertdiag}. If we view the self energy as the contribution to $\Gamma^2$ of $\Gamma^2(\vec k,\omega) = \Gamma_0^2(\vec k,\omega) - \Sigma(\vec k,\omega) + \text{counter terms}$, then from the diagrams in the figure, we see that the second renormalization condition becomes
\begin{equation}
\therefore \frac{d\Gamma^2}{d\omega} = 1 \Rightarrow \frac{d\Sigma}{d\omega}{((0,0),M)}+\delta_Z = 0
\end{equation}
So, to understand $Z$ we need only consider $d\Sigma/d\omega$. Since the contributions of the last two counterterm diagrams and the diagram with the ``$q=0$" bubble in Fig. \ref{fig:Pertdiag} are independent of $\omega$, these terms are then unimportant here. Hence we obtain
\begin{eqnarray}
&& \frac{d\Sigma(\vect{k},\omega)}{d\omega} = \frac{d}{d\omega}\int \frac{d^2qdq_0}{(2\pi)^3} \frac{16}{N|q|} \\
&& \times \frac{1+\frac{Z^2}{m^2_b}(2 k_x-q_x)^2+
\frac{Z^2}{m^2_b}(2 k_y-q_y)^2}{\big[(\omega -q_0)-((\vect{k}-\vect{q})^2/2m + |\mu|) + i\epsilon) \big]}
\label{eq:self_energy} \nonumber
\end{eqnarray}
Taking the derivative, we then obtain 
\begin{eqnarray}
&& \frac{d\Sigma}{d\omega}{((0,0),M)}  = -\frac{16}{N}\int \frac{d^3q}{(2\pi)^3} \frac{1}{|q|} \\
&&  \times \frac{1+\frac{Z^2}{m^2_b}(q_x)^2 + \frac{Z^2}{m^2_b}(q_y)^2}
{\big[q_0 - i\epsilon - M  + |\vect{q}|^2/2m + |\mu|) ) \big]^2} \nonumber
\end{eqnarray}

To understand this integral, we need the following identity, whose origin lies in the branch cut singularity:
\begin{eqnarray}
&& \lim_{\epsilon\to0} \int^\infty_{-\infty} \frac{dx}{(x-(a+i\epsilon)))^n\sqrt{x^2+b^2}} \\
&& = 2 \int^\infty_0 \frac{dy}{(i(b+y)-a)^n \sqrt{y^2+2by}},\quad b > 0 \nonumber
\label{eq:identity}
\end{eqnarray}
Using this, we then obtain
\begin{eqnarray}
&& \frac{d\Sigma}{d\omega}{((0,0),M)} = A \int^\infty_{-\infty} dq_x dq_y \int^\infty_0 dy \\
&& \times \frac{1+\frac{Z^2}{m^2_b}(q_x)^2 + \frac{Z^2}{m^2_b}(q_y)^2}
{(i(y+|\vect{q}|) - M + |\vect{q}|^2/2m + |\mu|)^2} \nonumber\\
&& \times \frac{1}{\sqrt{y^2+2|\vect{q}|y}} \nonumber
\end{eqnarray}
Finally, to understand the behavior of the above integral in the infra-red, we introduce
an UV cut-off $\Lambda$ for the integral and express everything
with the scale of energy in units of this UV cut-off $\Lambda$.
\begin{eqnarray}
&& \ti{q}_i=\frac{q_i}{\Lambda} , \ti{y}=\frac{y}{\Lambda} , \ti{M}=\frac{M}{\Lambda},
|\ti{\mu}|=\frac{|\mu|}{\Lambda},\ti{m}=\frac{m}{\Lambda}.
\end{eqnarray}	
In terms of these rescaled variables, we then obtain
\begin{eqnarray}
&& \frac{d\Sigma}{d\omega}{((0,0),\ti{M})} = A \int^1_{-1} d\ti{q}_x d\ti{q}_y \int^1_0 d\ti{y} \\
&& \times \frac{1+\frac{Z^2}{m^2_b}(\ti{q}_x)^2 + \frac{Z^2}{m^2_b}(\ti{q}_y)^2}
{(i(\ti{y}+|\vect{\ti{q}}|) - \ti{M} + |\vect{\ti{q}}|^2/2\ti{m} + |\ti{\mu}|)^2} \nonumber\\
&& \times \frac{1}{\sqrt{\ti{y}^2+2|\vect{\ti{q}}|\ti{y}}} \nonumber
\end{eqnarray}
Consider the infra-red limit $\lim \ti{M} \rightarrow 0$, i.e. when the renormalization 
scale is in the infra-red. The self-energy integral simplifies to
\begin{eqnarray}
&& \frac{d\Sigma}{d\omega}{((0,0),\ti{M})} = A \int^1_{-1} d\ti{q}_x d\ti{q}_y \int^1_0 d\ti{y} \\
&& \times \frac{1 + \frac{Z^2}{m^2_b}|\vect{\ti{q}}|^2}
{(i(\ti{y}+|\vect{\ti{q}}|) + |\vect{\ti{q}}|^2/2\ti{m}  + |\ti{\mu}|)^2} \nonumber\\
&& \times \frac{1}{\sqrt{\ti{y}^2+2|\vect{\ti{q}}|\ti{y}}} \nonumber
\end{eqnarray}
The only singularity in the integrand comes from $\ti{y} \rightarrow 0$, $|\vect{\ti{q}}| \rightarrow 0$
because of $\sqrt{\ti{y}^2+2|\vect{\ti{q}}|\ti{y}}$. The presence of $|\ti{\mu}|$ automatically
regulates the other denominator. Let's say we do the $\ti{y}$ integration first. Since, $\ti{y}$
will dominate over $\ti{y}^2$ when $\ti{y} \rightarrow 0$, therefore $\sqrt{\ti{y}^2+2|\vect{\ti{q}}|\ti{y}}
\approx  \sqrt{2|\vect{\ti{q}}|\ti{y}}$. As an integral over $\ti{y}$, this is an integrable
singularity. Similarly as an integral over (radial) $|\vect{\ti{q}}|$, the $\sqrt{2|\vect{\ti{q}}|\ti{y}}$
is an integrable singularity.

Thus, in the infra-red limit  $ \ti{M} \rightarrow 0$, $d\Sigma/d\omega$ is finite and hence can't have a logarithmic contribution($\log{\ti{M}} = \log{M/\Lambda} $). Through similar arguments, we can show that $\Sigma(\omega)$ is finite too. Also, the integral coming from the third renormalization condition in Eqn. \ref{eq:RG_cond} which determined $\delta_m$ will be finite as well. We mention this because in integral for $d\Sigma/d\omega$, we have the term $Z^2/m^2_b$ in the
numerator and $m_b$ depends on $m$ and $\delta_m$. But the finiteness of `$\delta_m$' integral guarantees us that there wouldn't be any logarithmic contribution to $d\Sigma/d\omega$ because of the presence of the $Z^2/m^2_b$ term. Thus, our earlier conclusion - that $d\Sigma/d\omega$ has no logarithmic divergence - remains valid. In absence of logarithmic divergences the bosonic
field won't acquire an anomalous dimension \cite{FieldTheory}, and thus $Z=1$. 

\subsection{Large-N perturbative calculations for an electron}

In the previous section we showed that large-N corrections to the
holons are finite and thus the $1/N$ holon
propagator is just a  renormalized form of the mean-field
holon propagator. As has been shown before by others, the $1/N$ spinon propagator does get a logarithmic correction.
In this section, we work out the consequences of these two
statements on to the electron propagator made of the a holon and a
spinon that would be measured in an ARPES experiment. 

\begin{figure}
\includegraphics[width=0.95\linewidth]{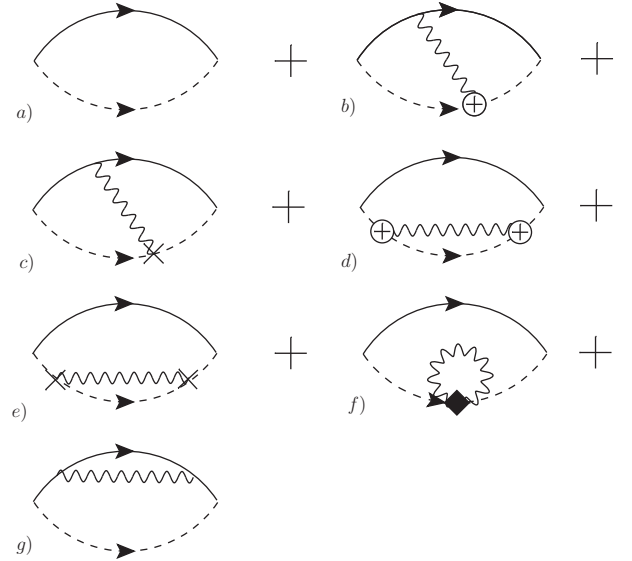}
\caption{The terms in the perturbation series for the Hole
propagator up to $O[1/N]$}
\label{fig:Hole_LargeN}
\end{figure}

The Feynman diagrams for the electron propagator to $O[1/N]$ are shown 
in Fig. \ref{fig:Hole_LargeN}. We are interested in finding out if the hole propagrator 
recieves any logarithmic corrections due to the gauge fluctuations due 
to divergent diagrams. Let us first look at Fig. \ref{fig:Hole_LargeN} a)
which is the continuum limit of the mean-field electron propagator 
$A_{mean}(\vec q,\omega)$ studied in Sec. III and shown to be linear
in $\omega$ through an analysis of mean-field band structure of spinons
and holons and through a mean-field scaling argument. For our low-energy 
continuum theory (we use the same conventions
for the spinon propagators as in Ref. \onlinecite{Hermele1}), this diagram
is equal to
\begin{equation}
  {\bf A}_{a}(\vec q,\omega) = \int d\omega_1 d\vect{q}_1 {\bf S}(\vec q_1,\omega_1)G(\vec q-\vec q_1,\omega-\omega_1)
\end{equation}
where ${\bf A}_{a}$ is matrix spectral function that enters Eq. \ref{eq:ConSpec} 
and is related to the measured spectral function through matrix elements,
\begin{equation}
 {\bf S}(\vec q,\omega)=\frac{1}{\gamma^0 (\omega + i \epsilon sgn(\omega))
+ \vect{\gamma}\cdot\vect{q}}
\end{equation} 
and
\begin{equation}
 G(\vec q,\omega) = \frac{1}{(\omega -\omega_1) + i\epsilon - \frac{|\vect{q}-\vect{q}_1|^2}{2m_b}-|\mu_b|}
\end{equation}
is the boson propagator entering Eq. \eqref{eq:bpropagator}. Evaluating the frequency integral we arrive at
%\onecolumngrid
%\begin{eqnarray}
%\label{eq:diagA}
%& = & \int d\omega_1 d\vect{q}_1 \frac{1}{[\gamma^0 (\omega_1 + i \epsilon sgn(\omega_1))
%+ \vect{\gamma}\cdot\vect{q}][(\omega -\omega_1) + i\epsilon - \frac{|\vect{q}-\vect{q}_1|^2}{2m}-|\mu|]} 
%\nonumber  \\
%& = & \int d\omega_1 d\vect{q}_1 \frac{\gamma^0 (\omega_1 + i \epsilon sgn(\omega_1))
%+ \vect{\gamma}\cdot\vect{q}_1}{[\omega^2_1+|\vect{q}_1|^2]
%[(\omega -\omega_1) + i\epsilon - \frac{|\vect{q}-\vect{q}_1|^2}{2m}-|\mu|]} \nonumber \\ 
%& & \text{doing } \omega_1 \text{-integral as a contour integral by closing on lower-half plane} \nonumber \\
\begin{equation}\label{eq:diagA}
{\bf A}_a(\vec q,\omega) =  \pi \int d\vect{q}_1 \frac{i \gamma^0 - \vect{\gamma}\cdot\unit{q}_1}{\omega+i|\vect{q}_1|
-\frac{|\vect{q}-\vect{q}_1|^2}{2m_b}-|\mu_b|}
\end{equation}
%\twocolumngrid
In the infrared limit at the Dirac point $\omega \rightarrow 0$, $\vect{q}=(0,0)$,
the integrand in Eq. \ref{eq:diagA} has a finite denominator, thus there is no divergence.
This is as we would have expected of the mean-field hole propagator and the leading
term (in $\omega$) of the (finite) integral is linear in $\omega$ (as concluded from 
other arguments in Sec III). Since in the previous subsection we showed 
that the holon recieves only finite corrections at $1/N$, the diagrams
Fig. \ref{fig:Hole_LargeN} d), e) and f) 
%(with $O[1/N]$ corrected holon propagator which is just a renormalized form of the mean-field propagator) 
are also finite as can be shown in a similar manner as the evaluation of 
the mean-field diagram above. Hence, diagrams a), d), e) and f) are all finite 
diagrams as far as divergences are considered and are consistent with the mean field scaling.

We proceed to look at the diagram Fig. \ref{fig:Hole_LargeN} b) and c) where a photon 
is exchanged between the holon and the spinon. This diagram equals to
\begin{multline}
{\bf A_b}(\vec q,\omega)=\!\!\int d\omega_1 d\vect{q}_1 d\omega_2 d\vect{q}_2\bigg[\\
{\bf S}(\vec q_1,\omega_1){\bf S}(\vec q_1-\vec q_2,\omega_1-\omega_2)
\\
 \times G(\vec q-\vec q_1,\omega-\omega_1)G(\vec q-\vec q_1+\vec q_2,\omega-\omega_1+\omega_2) \\
  \gamma^\mu D_{\mu\nu}(\omega_2,\vec q_2) [V^\nu(\vec q - \vec q_1) + V^\nu(\vec q - \vec q_1 + \vec q_2)]\bigg]
\end{multline}
where $D_{\mu\nu}(\vec q,\omega)$ is the photon propagator 
of Eq. \eqref{eq:photon_propagator} and $V^\mu(\vec k)$ is 
defined in Eqs. \eqref{eq:bosontimevertex} and \eqref{eq:bosonspacevertex}.
%\begin{eqnarray}
%\label{eq:diagB}
%& = & \int d\omega_1 d\vect{q}_1 d\omega_2 d\vect{q}_2
%\left [\frac{\gamma^0 \omega_1+ \vect{\gamma}\cdot\vect{q}_1}{\omega^2_1+|\vect{q}_1|^2} \right]
%\left[ \frac{\gamma^0 (\omega_1 - \omega_2) + \vect{\gamma}\cdot(\vect{q}_1-\vect{q}_2}
%{(\omega_1-\omega_2)^2+|\vect{q}_1-\vect{q}_2|^2} \right]
%\left[ \frac{16}{N\sqrt{\omega^2_2+|\vect{q}_2|^2}}\right] \times
%\nonumber \\
%& & \times
%\left[ \frac{1}{(\omega -\omega_1) + i\epsilon - \frac{|\vect{q}-\vect{q}_1|^2}{2m}-|\mu|} \right]
%\left[ \frac{1}{(\omega -\omega_1+ \omega_2) + i\epsilon - \frac{|\vect{q}-\vect{q}_1+\vect{q}_2|^2}{2m}-|\mu|} \right]
%\nonumber \\
Doing the $\omega_1$-integral as a contour integral by closing on lower-half plane %[{\bf add "Vertex corrections!"}]
we arrive at
%\onecolumngrid
\begin{multline}
 =  \frac{16 \pi}{N} \int d\vect{q}_1 d\omega_2 d\vect{q}_2
 \bigg[ \frac{i \gamma^0 - \vect{\gamma}\cdot\unit{q}_1}{\omega+i|\vect{q}_1|
-\frac{|\vect{q}-\vect{q}_1|^2}{2m}-|\mu|} \\
\times \frac{\gamma^0 (-i|\vect{q}_1| - \omega_2) + \vect{\gamma}\cdot(\vect{q}_1-\vect{q}_2)}
{(\omega +i|\vect{q}_1|+ \omega_2) + i\epsilon - \frac{|\vect{q}-\vect{q}_1+\vect{q}_2|^2}{2m}-|\mu|}  \\
\times \frac{\gamma^\mu [V^\mu(\vec q - \vec q_1) + V^\mu(\vec q - \vec q_1 + \vec q_2)]}
{\sqrt{\omega^2_2+|\vect{q}_2|^2}} \bigg]
\end{multline}
and using Eq. \ref{eq:identity}  to do the $\omega_2$ integral this further reduces to
\begin{multline}
 =  \frac{32 \pi}{N} \int d\vect{q}_1 d\vect{q}_2
\bigg[ \frac{i \gamma^0 - \vect{\gamma}\cdot\unit{q}_1}{\omega+i|\vect{q}_1|
-\frac{|\vect{q}-\vect{q}_1|^2}{2m}-|\mu|}\\\times
 \int^\Lambda_0 dy\frac{\gamma^0 (-i|\vect{q}_1| - i(|\vect{q}_2|+y)) + \vect{\gamma}\cdot(\vect{q}_1-\vect{q}_2)}
{(\omega +i|\vect{q}_1|+ i(|\vect{q}_2|+y)) - \frac{|\vect{q}-\vect{q}_1+\vect{q}_2|^2}{2m}-|\mu|}\\
\times \gamma^\mu [V^\mu(\vec q - \vec q_1) + V^\mu(\vec q - \vec q_1 + \vec q_2)]
\frac{1}{\sqrt{y^2+2y|\vect{q}_2|}} \bigg]
\end{multline}
%\twocolumngrid
Looking at $\omega \rightarrow 0$, $\vect{q}=(0,0)$ again 
(the infrared limit at the Dirac point), we see that the only 
singular piece in the integrand is $\frac{1}{\sqrt{y^2+2y|\vect{q}_2|}}$ 
which (as argued in Sec III b) is an integrable singularity. Thus, the 
diagrams Fig. \ref{fig:Hole_LargeN} b) and c) give only finite corrections to the hole propagator.

\begin{figure}
\psfrag{Imom}{$\Im[\omega_1]$}
\psfrag{Reom}{$\Re[\omega_1]$}
\psfrag{iq}{$i |\vect{q}_1|$}
\psfrag{miq}{$- i |\vect{q}_1|$}
\includegraphics[width=0.95\linewidth]{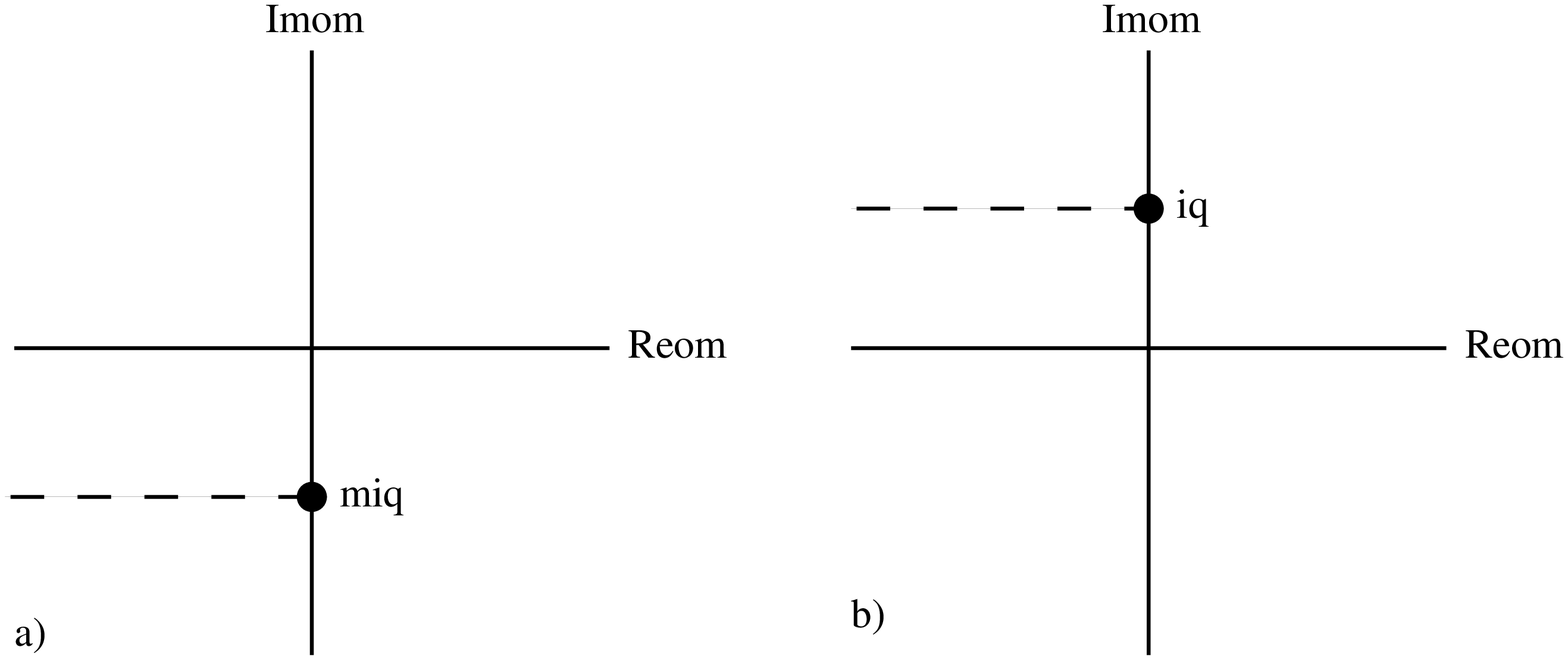}
\caption{We show the branch cuts in a) of $\log(\frac{\omega_1+i|\vect{q}_1|}{\Lambda})$ 
and in b)} of  $\log(\frac{\omega_1-i|\vect{q}_1|}{\Lambda})$
\label{fig:Branch_Cut}
\end{figure}

We finally come to the diagram Fig. \ref{fig:Hole_LargeN} g), where the loop
on the spinon propagator has been shown to give a logarithmic correction to the
spinon propagator. We want to know what is the effect of this logarithmic
correction on the electron propagator. This diagram equals
\onecolumngrid
\begin{eqnarray}
\label{eq:diagD_1}
& = & \frac{8}{3\pi^2N} \int d\omega_1 d\vect{q}_1
\left[ \frac{(\gamma^0 \omega_1+ \vect{\gamma}\cdot\vect{q}_1) \log(\sqrt{\omega^2_1+|\vect{q}_1|^2}/\Lambda)}
{\omega^2_1+|\vect{q}_1|^2} \right]
\left[ \frac{1}{(\omega -\omega_1) + i\epsilon - \frac{|\vect{q}-\vect{q}_1|^2}{2m}-|\mu|} \right]
\end{eqnarray}
\twocolumngrid
\noindent where the logarithm is due to photon-spinon loop (see Eq. C9 of Ref. \onlinecite{Hermele1}).
We can express $\log(\sqrt{\omega^2_1+|\vect{q}_1|^2}/\Lambda)$ as
$(1/2) (\log((\omega_1+i\vect{q}_1/\Lambda) + \log((\omega_1-i\vect{q}_1)/\Lambda))$.
As a function of (complex) $\omega_1$, we show the branch cut structure due to the
logarithms in Fig. \ref{fig:Branch_Cut}. Thus we will do the $\omega_1$ integral
as a contour integral keeping in mind that for the $\log((\omega_1+i\vect{q}_1/\Lambda)$
term, we close the contour on the upper-half plane, while for 
$\log((\omega_1-i\vect{q}_1/\Lambda)$, we close the contour on the lower-half
plane. By capturing residues we get that the diagram equals
%\onecolumngrid
\begin{eqnarray}
\label{eq:diagD_2}
& = & A + B + C \text{   ... where} \nonumber \\
A & = & \frac{4}{3\pi N} \int d\vect{q}_1 \frac{(-i \gamma^0 + \vect{\gamma}\cdot\unit{q}_1)\log(-\frac{2i|\vect{q}_1|}{\Lambda})}
{\omega+i|\vect{q}_1|-\frac{|\vect{q}-\vect{q}_1|^2}{2m}-|\mu|} \nonumber \\
B & = & \frac{4}{3\pi N} \int d\vect{q}_1 \frac{(i \gamma^0 + \vect{\gamma}\cdot\unit{q}_1)\log(\frac{2i|\vect{q}_1|}{\Lambda})}
{\omega-i|\vect{q}_1|-\frac{|\vect{q}-\vect{q}_1|^2}{2m}-|\mu|} \nonumber \\
C & =  & - \frac{8i}{3\pi N} \int d\vect{q}_1
\frac{(\gamma^0(\omega-\frac{|\vect{q}-\vect{q}_1|^2}{2m}-|\mu|)+\vect{\gamma}\cdot\vect{q}_1)}
{(\omega-\frac{|\vect{q}-\vect{q}_1|^2}{2m}-|\mu|)^2+|\vect{q}_1|^2} \nonumber \\
& &\times\log(\frac{\omega+i|\vect{q}_1|-\frac{|\vect{q}-\vect{q}_1|^2}{2m}-|\mu|}{\Lambda})
\end{eqnarray}
%\twocolumngrid
In the infrared limit at the Dirac point (i.e. $\omega \rightarrow 0$, $\vect{q}=(0,0)$),
the above simplifies to
%\onecolumngrid
\begin{eqnarray}
\label{eq:diagD_3}
& = & A + B + C \text{   ... where} \nonumber \\
A & = & \frac{4}{3\pi N} \int d\vect{q}_1 \frac{(-i \gamma^0 + \vect{\gamma}\cdot\unit{q}_1)
(\log(\frac{2|\vect{q}_1|}{\Lambda})-\frac{i\pi}{2})}
{\omega+i|\vect{q}_1|-\frac{|\vect{q}_1|^2}{2m}-|\mu|} \nonumber \\
B & = & \frac{4}{3\pi N} \int d\vect{q}_1 \frac{(i \gamma^0 + \vect{\gamma}\cdot\unit{q}_1)
(\log(\frac{2|\vect{q}_1|}{\Lambda})+\frac{i\pi}{2})}
{\omega-i|\vect{q}_1|-\frac{|\vect{q}_1|^2}{2m}-|\mu|} \nonumber \\
C & =  & - \frac{8i}{3\pi N} \int d\vect{q}_1
\frac{(\gamma^0(\omega-\frac{|\vect{q}_1|^2}{2m}-|\mu|)+\vect{\gamma}\cdot\vect{q}_1)}
{(\omega-\frac{|\vect{q}_1|^2}{2m}-|\mu|)^2+|\vect{q}_1|^2} \nonumber \\
& &\times\log(\frac{\omega+i|\vect{q}_1|-\frac{|\vect{q}_1|^2}{2m}-|\mu|}{\Lambda})
\end{eqnarray}
%\twocolumngrid
The $\vect{\gamma}\cdot\unit{q}_1$ and $\vect{\gamma}\cdot\vect{q}_1$ terms are odd under inversion and hence
upon integration equal zero. $A+B$ is finite. This can be seen by expanding the denominators of $A$ and $B$
in $|\vect{q}_1|$. At lowest order for $|\vect{q}_1|$, 
the two (singular) logarithmic contributions cancel each other out. Higher orders in $|\vect{q}_1|$ 
don't have singular integrands as $|\vect{q}_1|^n \log(|\vect{q}_1|) \rightarrow 0$ as
$|\vect{q}_1| \rightarrow 0$ for $n \geq 1$.

We can write the logarithm in $C$ as $\log(\frac{\sqrt{(\omega-\frac{|\vect{q}_1|^2}{2m}-|\mu|)^2
+|\vect{q}_1|^2}}{\Lambda}) + i \arctan(\frac{|\vect{q}_1|}{\omega-\frac{|\vect{q}_1|^2}{2m}-|\mu|})$.
We focus on the imaginary part of $C$ since it is the most singular part. We can write this as
%\onecolumngrid
\begin{eqnarray}
\label{eq:diagD_4}
\Im[C] & = & - \frac{4}{3\pi N} \int d\vect{q}_1
\frac{\gamma^0(\omega-\frac{|\vect{q}_1|^2}{2m}-|\mu|)}
{(\omega-\frac{|\vect{q}_1|^2}{2m}-|\mu|)^2+|\vect{q}_1|^2}\nonumber \\
& &\times\log(\frac{(\omega-\frac{|\vect{q}_1|^2}{2m}-|\mu|)^2+|\vect{q}_1|^2}{\Lambda^2}) \nonumber \\
& = & - \frac{8}{3 N} \int^\Lambda_0 m dy 
\frac{\gamma^0(\omega-|\mu|-y)
\log(\frac{(\omega-|\mu|-y)^2+2my}{\Lambda^2})}
{(\omega-|\mu|-y)^2+2my} \nonumber \\
\end{eqnarray}
%\twocolumngrid
and we see that the integrand is finite everywhere on the integration domain (with the UV cut-off).
Therefore the integral cannot have any singular/logarithmic term. Thus we conclude that 
$O[1/N]$ corrections to the hole propagator due to the gauge fluctuations are finite.
Therefore the $O[1/N]$ scaling of the ARPES spectral function is the \emph{same}
as the mean-field scaling.

%\section{On $\vect{k}$ dependence}

%\section{Conclusion and Dicussion}

\section{Discussion}
%\textbf{ Discuss each result and its importance for the broader picture}
% A) Linear deendence on \omega in A(Q_f-Q_b,\omega) only at specific momenta. 
% B) Specialness of Q_f-Q_b is also a simple experimental observable.
%     --> This is not generic
%     -->  
% C) Renormalization of spinon but not holon 
%      --> dramatic spin charge separation
%      --> 

The key prediction of our work is that the ARPES spectral function for the putative spin liquid,
Herbertsmithite, will show a linear low-energy behavior if the ground state has Dirac
Spinons. This prediction is expected to be unique to the Dirac Spin Liquid as the Dirac point-like
Fermi ``surface" is not generic. Thus, other spin liquid ground states might generically be
expected to have different energy dependences. Recently, NMR experiments \cite{LeeSingleCrystal1}
 were performed on single crystal
Herbersmithite and the authors reported that their results are not in contradiction with a
DSL. This makes the case for an ARPES experiment on Herbertsmithite even more exciting since,
as stressed before,
our prediction is a falsifiable prediction.

Another unique feature of our prediction is based on the fact that DSL is a high symmetry state in momentum
space. Thus, the linear energy dependence will be most clearly observable at very
specific momenta (i.e., momenta that connect these high symmetry points in the Brillouin 
Zone coming from the Dirac points for spinons and the dispersion minima for the holons
$Q^m_f - Q^n_b$)).
Observation of these momenta would be another strong evidence for a DSL.

We found that the holons in the Dirac spin liquid
get trivial renormalizations from the gauge field 
fluctuations and remain as free bosons.
On the other hand, as has been shown in earlier literature,
the Dirac spinons do get nontrivial logarithmic renormalization
from the gauge field fluctuations. This feature
points to a dramatic spin-charge separation in the Dirac spin liquid.

%\subsection{Discussion}
%\textbf{ If our proposal is successful and the spinon is found, ...}
In this paper, we have discussed how ARPES can provide an avenue to prove
the existence of spinons in Kagome Antiferromagnet. Existence of spinons would raise
two obvious questions : 1) How would one detect the corresponding charged 
holons ?, and 2)
How would one detect the photons corresponding to the U(1) gauge field ?

Detection of holons can be envisaged perhaps through a transport experiment. Since the 
holons are charged bosons, we can expect to probe their existence in
transport measurements. Since the low energy holons occur at particular non-zero
momenta in the zone, there might be a non-trivial angular dependence of 
conductance in a sample which can be compared against theoretical
predictions. Detection of the
``photon" is probably a harder task. These photons will of course contribute
to the specific heat along with spinons and one will have to separate
the two contributions. Thermal conductivity measurements can similarly
get contributions from the photons.

%\textbf{ End this with a discussion of the "universe".}
In conclusion, frustrated quantum magnets offer an 
universe of interesting objects including the one
we focused on in this study : Dirac spinons, U(1) photons and non-interacting bosons
built out of a Dirac spin liquid state. Herbertsmithite provides with an ideal 
candidate material to explore this universe and find these exotic objects. 
Finding the Spinon in Nature will certainly be a big step in the uncovering of that 
universe and a successful ARPES experiment on Herbertsmithite
might go a long way in this regard.

%The questions raised above are natural questions once given the existence of spinons. Thus, the search for spinons in the frustrated Kagome Antiferromagnet,Herbertsmithite, and also in other frustrated magnets assumes significance. A successful ARPES experiement might go a long way in settling questions in this regard.

\emph{Acknowledgements.} We thank Kyle Shen, Chris Henley and Siddharth Parmeswaran for
useful discussions. During the work, SP was supported by NSF grant DMR 1005466.


\begin{thebibliography}{99}
\bibitem{e2footnote}
Here we are supressing the $e$ in otherwise
$D_\mu = \partial_\mu - i e A_\mu(\vect{r},t)$, keeping in mind that $e^2 \sim 1/N$
ensures a perturbation series for the holon with $1/N$ as the small parameter
as is usual in large N expansions.
\end{thebibliography}
\end{document}